\documentclass[epj]{svjour}
\usepackage{graphics,epsfig}
\begin{document}
\title{Predictions  
for top quark spin correlations 
 at the Tevatron and the LHC at next-to-leading order in $\alpha_s$}
\author{A. Brandenburg
}                     
%
%
\institute{Institut f\"ur Theoretische Physik E, 
RWTH Aachen, D-52056 Aachen, Germany}
\date{Received: date / Revised version: date}
%
\abstract{
Predictions for 
angular distributions of top
quark decay products that are sensitive to $t\bar{t}$ spin
correlations are presented at next-to-leading order in 
$\alpha_s$ for the Tevatron and the LHC.
} 
\authorrunning{A. Brandenburg}
\titlerunning{Predictions for top quark spin correlations at the
  Tevatron and the LHC \ldots}
\maketitle
\section{Introduction}
\label{intro}
One of the striking features of the top quark is that due to 
its extremely short lifetime it cannot form hadronic bound states.
The strong interactions involved in the dynamics of top quark production
and decay are thus reliably described by perturbative QCD. 
The Standard Model main decay mode $t\to Wb$ is parity violating.
Therefore, the spin properties of top quarks are transferred
to its decay products without being diluted by hadronization,
and they become additional observables to study the interactions of the
top quark. Spin observables are useful to provide 
constraints on fundamental parameters of the SM
(e.g. $V_{tb}$), to probe possible
new production mechanisms for top quarks, to 
search for non-standard couplings in top quark decays, and 
to test discrete symmetries 
(like CP). At hadron colliders, the top quarks are produced 
predominantly in pairs via pure QCD processes. Since QCD preserves
parity, the top quarks and antiquarks produced
in the parton processes $q\bar{q}\to t\bar{t},gg\to t\bar{t}$ 
are unpolarized at leading order in $\alpha_s$. 
At next-to-leading order (NLO), 
absorptive parts in the scattering amplitudes of the above parton
reactions lead to a nonzero polarization of the top quarks and antiquarks 
perpendicular to the event plane. It is a quite small
effect, of the order of a percent \cite{ber96,dha96}.
Much larger are the correlations of the spins of top quark and antiquark:
In fact, provided one chooses an adequate spin basis, the correlations
are of the order of 1. They can be studied 
by measuring double differential angular distributions of
top quark decay products both at the Tevatron 
and at the LHC. In this talk,  results for  
these distributions  at NLO in the strong
coupling $\alpha_s$ will be discussed.\vfil
\section{Theoretical framework}
\label{theory}
We consider here  the following processes:
\begin{equation}
  h_1h_2\to t\bar{t}+X\to
  \left\{\begin{array}{lcc}
      \ell^+\ell '^{-} &+& X\\ 
      \ell^{\pm}j_{\bar{t}(t)} &+& X\\ 
      j_t j_{\bar{t}}&+&X
    \end{array} \right.,
  \label{reac1}
\end{equation}
where $h_{1,2}=p,\bar{p}$; $\ell=e,\mu,\tau$, and 
$j_t\ (j_{\bar{t}})$ denote  
jets originating from hadronic $t$ ($\bar{t}$) decays. 
An observable that 
is intimately related to the $t\bar{t}$ spin correlations
in the above reactions  is the double differential angular
distribution of the top decay products, e.g. for the dilepton channel:
\begin{equation}\label{dist}
\frac{1}{\sigma}\frac{d^2\sigma}{d\cos\theta_+d\cos\theta_-}
=\frac{1}{4}(1-C\cos\theta_+\cos\theta_-).
\end{equation}
In (\ref{dist}), $\theta_\pm$ are the angles between the $\ell^{\pm}$
direction of flight in the $t(\bar{t})$ rest frame (defined by a rotation-free
boost from the parton c.m.s.) with respect to
axes $\hat{\bf a}$ ($\hat{\bf b}$) which will be specified below. 
For our choices of 
$\hat{\bf a}$, $\hat{\bf b}$, 
terms linear in $\cos\theta_\pm$ are absent  
in (\ref{dist}) due to parity invariance of QCD.  

The calculation of the distribution (\ref{dist}) at 
NLO QCD simplifies 
enormously in the leading pole
approximation (LPA), which amounts to expanding the full
amplitudes for (\ref{reac1}) around the complex poles of the $t$ and
$\bar{t}$ propagators. Only the leading pole terms are kept in this 
expansion, i.e. one neglects terms of order $\Gamma_t/m_t\approx 1\%$.
Within the LPA, the radiative corrections can be classified into
factorizable and non-factorizable \cite{beenakker} contributions.
The impact of the non-fac\-torizable  contributions
will be discussed elsewhere \cite{bbsu03}. 
Here we consider only the factorizable corrections.
For these the coefficient in the distribution (\ref{dist}) factorizes:
\begin{equation}\label{c}
C = \kappa_+\kappa_-D.
\end{equation}
In (\ref{c}), $D$ is the $t\bar{t}$ double spin asymmetry
\begin{equation}
\label{d}
D=\frac{N(\uparrow\uparrow)+N(\downarrow\downarrow)-N(\uparrow\downarrow)
-N(\downarrow\uparrow)}{N(\uparrow\uparrow)
+N(\downarrow\downarrow)+N(\uparrow\downarrow)+N(\downarrow\uparrow)},
\end{equation}
where $N(\uparrow\uparrow)$ denotes the number of $t\bar{t}$ pairs
with $t$ ($\bar{t}$) spin parallel to the reference axis $\hat{\bf a}$ 
($\hat{\bf b}$) etc. The directions $\hat{\bf a}$ and $\hat{\bf b}$  
can thus be identified with the
spin quantization axes for $t$ and $\bar{t}$, and $D$ directly 
reflects the strength of the correlation between the 
$t$ and $\bar{t}$ spins for the chosen axes. 
In fact, a simple calculation shows
\begin{equation}
D={\rm corr}(\hat{\bf a}\cdot {\bf S}_t,\hat{\bf b}\cdot {\bf S}_{\bar{t}}),
\end{equation}
where ${\bf S}_{t(\bar{t})}$ denotes the spin operator of the top quark 
(antiquark) and ${\rm corr}(O_1,O_2)$ is the correlation of two
observables defined in the standard way. 

The prefactor $\kappa_\pm$ in (\ref{c}) is the spin analysing power
of the charged lepton in the decay 
$t(\bar{t})\to b(\bar{b})\ell^\pm\nu (\bar{\nu})$ 
defined by the decay distribution
\begin{equation}\label{decay}
\frac{1}{\Gamma}\frac{d\Gamma}{d\cos\vartheta_{\pm}}=\frac{1\pm\kappa_\pm
\cos\vartheta_\pm}{2},
\end{equation}
where $\vartheta_\pm$ is the angle between the $t$ $(\bar{t})$ spin
and the $\ell^\pm$ direction of flight.
It is clear that the value of $\kappa_{\pm}$ is crucial for the 
experimental determination of the $t\bar{t}$ spin correlations.
From \cite{cjk} we obtain
\begin{eqnarray} 
\kappa_+=\kappa_-=1-0.015\alpha_s,
\end{eqnarray}
which means that the charged lepton serves as a perfect analyser 
of the top quark spin. 
For hadronic decays $t\to bq\bar{q}'$ one has a decay distribution analogous
to (\ref{decay}) (and a double angular distribution
analogous to (\ref{dist})), and  
the spin analysing power of jets can be defined.
To order $\alpha_s$ \cite{bsu} (and using $\alpha_s(m_t)=0.108$), 
\begin{eqnarray} \label{kappa}
\kappa_b=-0.408\times(1-0.340\alpha_s)=-0.393,\\
\kappa_j=+0.510\times(1-0.654\alpha_s)=+0.474,
\end{eqnarray}   
where $\kappa_j$ is the analysing power of the least energetic non-b-quark
jet. 
Note that the loss of analysing power 
using hadronic final states is 
overcompensated by the gain
in statistics and in efficiency to reconstruct the $t$ ($\bar{t}$) rest frames.

To compute $D$  at NLO QCD, the differential
cross sections for the following parton processes are needed to order
$\alpha_s^3$, 
where the full
information on the $t$ and $\bar{t}$ spins has to be kept:  
\begin{eqnarray}\label{parton} 
q\bar{q}\to t\bar{t},\ t\bar{t}g;\ 
gg\to t\bar{t},\  t\bar{t}g;\  
q(\bar{q})g\to t\bar{t}q(\bar{q}).
\end{eqnarray}
Results  at NLO QCD for the $\overline{{\mbox{MS}}}$ 
subtracted parton cross sections
$\hat{\sigma}$ for the above processes
with $t\bar{t}$ spins summed over have been obtained in
\cite{nason,been}, 
while 
\begin{equation}
\hat{\sigma}\hat{D}= \hat{\sigma}(\uparrow\uparrow)+
\hat{\sigma}(\downarrow\downarrow)-\hat{\sigma}(\uparrow\downarrow)
-\hat{\sigma}(\downarrow\uparrow)
\end{equation}
has been computed for 
different spin quantization axes in \cite{us1,us2}.
This combination of spin-dependent parton cross sections can be written as
follows
\begin{eqnarray}
\hat{\sigma}\hat{D}=\frac{\alpha_s^2}{m_t^2}\left\{
g^{(0)}(\eta) +4\pi\alpha_s {\cal{G}}^{(1)}\right\},
\end{eqnarray}
with
\begin{eqnarray}
{\cal{G}}^{(1)}=g^{(1)}(\eta)+
\tilde{g}^{(1)}(\eta)\ln\left(\frac{\mu^2}{m_t^2}\right),
\end{eqnarray}
where $\eta=\hat{s}/(4m_t^2)-1$, and we use a common scale $\mu$ for 
the renormalization and factorization scale. As an example, Fig. 1 shows, 
for the parton process $gg\to t\bar{t}(g)$,   
the three functions that determine  $\hat{\sigma}\hat{D}$ 
for the beam basis, i.e. both $\hat{\bf a}$ and $\hat{\bf b}$ 
are chosen to be along one of the hadron beams in the laboratory frame.
\begin{figure}[t]
\unitlength1.0cm
\begin{center}
\begin{picture}(7.5,7.5)
\put(0,0.5){\psfig{figure=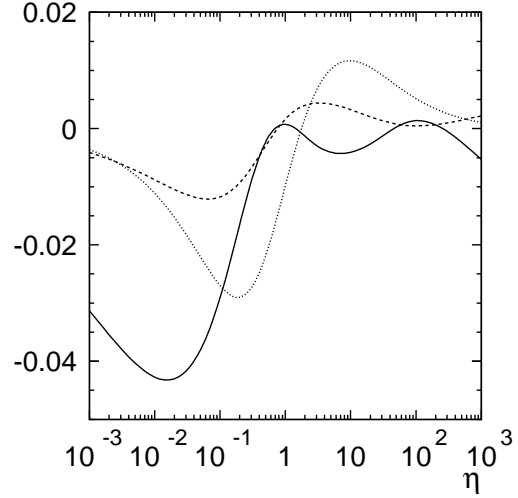,width=7.5cm,height=7.5cm}}
\end{picture}
\\[-0.5cm]
\caption{Dimensionless scaling functions $g^{(0)}(\eta)$
(full), $g^{(1)}(\eta)$ (dotted), and
${\tilde g}^{(1)}(\eta)$ (dashed) that determine 
$\hat{\sigma}\hat{D}$ in the beam basis 
for the $gg$ initial state.}\label{fig:siggg}
\end{center}
\end{figure}
Apart from the beam basis we also consider the helicity basis, where
$\hat{\bf a}$ ($\hat{\bf b}$) is chosen to be
the $t$ ($\bar{t}$) direction of flight in the parton c.m.s.
, and the so-called off-diagonal
basis \cite{off}, which is defined by the requirement that 
$\hat{\sigma}(\uparrow\downarrow)$ $=\hat{\sigma}(\downarrow\uparrow)=0$
for the process $q\bar{q}\to t\bar{t}$ at tree level.
\section{Predictions for the Tevatron and the LHC}
\label{results}
In Table 1 we list our predictions \cite{bbsu03,us3} 
for the spin correlation coefficient $C$ in the double 
differential distribution (\ref{dist}) at the Tevatron
for the three 
different choices of spin quantization axes discussed above. 
\begin{table}[h]\begin{center} \renewcommand{\arraystretch}{1.25}
\caption{Results for the spin correlation coefficient $C$ of 
the distribution (\ref{dist}) at LO and
NLO for 
$p\bar{p}$ collisions at $\sqrt{s}=1.96$ TeV for different $t\bar{t}$
decay modes. The PDFs CTEQ6L (LO) and CTEQ6M (NLO) of \cite{CTEQ6} 
were used, and $\mu=m_t=175$ GeV.}
\label{tab:Tevatron}       
\begin{tabular}{|ccccc|}  
\hline
          &    & dilepton    &lepton+jet  & jet+jet  \\ \hline
$C_{\rm hel}$ &LO  & $-$0.471     & $-$0.240  & $-$0.123 \\
          &NLO & $-$0.387     & $-$0.185  & $-$0.088 \\ \hline
$C_{\rm beam}$ &LO &  $\phantom{-}$0.928    &  $\phantom{-}$0.474  
&  $\phantom{-}$0.242 \\ 
           &NLO&  $\phantom{-}$0.801    &  $\phantom{-}$0.382  &  
$\phantom{-}$0.182\\ \hline
$C_{\rm off}$   &LO& $\phantom{-}$0.937    &  $\phantom{-}$0.478 &  
$\phantom{-}$0.244 \\ 
           &NLO& $\phantom{-}$0.808    &  $\phantom{-}$0.385 &  
$\phantom{-}$0.183\\ 
\hline
\end{tabular}\end{center}
\end{table} 
We use the CTEQ6L (LO) and CTEQ6M (NLO) parton distribution functions
(PDFs) \cite{CTEQ6}.
Numbers are given for the dilepton, lepton$+$jet and all-hadronic
decay mode of the $t\bar{t}$ pair, where in the latter two cases 
the least energetic non-$b$-quark jet was used as spin analyser. 
One sees that the spin correlations are largest in the
beam and off-diagonal basis. The QCD corrections reduce the LO results
for the coefficients $C$ by about 10\% to 30\%.

For the LHC it turns out that the spin correlations w.r.t. the 
beam and off-diagonal basis are quite small due to a cancellation 
of contributions from the $gg$ and $q\bar{q}$ initial states. 
Here, the helicity basis is
a good choice, and Table~2 lists our results for the $C$ coefficient
in that case. The QCD corrections are smaller  at the LHC
than at the Tevatron and vary between 1 and 10$\%$.
At both colliders the relative corrections 
$|(C_{\rm NLO}-C_{\rm LO})/C_{\rm LO}|$ 
are largest for the all-hadronic decay modes and smallest for the
dilepton decay modes.

\begin{table}[h]\begin{center} \renewcommand{\arraystretch}{1.25}
\caption{Results for  $C_{\rm hel}$ 
for $pp$ collisions at $\sqrt{s}=14$ TeV using the same PDFs and
parameters
as in Tab. 1. }
\label{tab:LHC}       
\begin{tabular}{|ccccc|}   
\hline
   &    & dilepton    &lepton+jet  & jet+jet  \\ \hline
$C_{\rm hel}$ &LO  & 0.319    &  0.163 &  0.083 \\
          &NLO & 0.322    &  0.156 &  0.076 \\ 
\hline
\end{tabular}
\end{center}
\end{table}
Table 3 shows the dependence of the NLO results 
on the choice of the PDFs for dilepton final states.
While the results for the CTEQ6 and MRST2002 \cite{MRST02} PDFs agree
at the percent level, the GRV98 \cite{GRV98} PDFs 
give significantly different results
at the Tevatron. This is 
largely due to the fact that the contributions from $q\bar{q}$ and $gg$ initial
states contribute to $C$ with opposite signs. This may offer the possibility
to constrain the quark and gluon content of the proton by a precise 
measurement of the double angular distribution (\ref{dist}).
\begin{table}[h]\begin{center} \renewcommand{\arraystretch}{1.25}
\caption{Spin correlation coefficients at NLO for different PDFs for
 $p\bar{p}$ at $\sqrt{s}=1.96$ TeV (upper part) and 
 $pp$ at $\sqrt{s}=14$ TeV (lower part) for dilepton final states.}
\label{tab:PDF}
\begin{tabular}{|cccc|}  \hline
\multicolumn{4}{|c|}{Tevatron} \\ \hline
              &  CTEQ6M  &MRST2002       &GRV98\\ \hline
$C_{\rm hel}$
          & $-$0.387    &  $-$0.384    & $-$0.328  \\  \hline
$C_{\rm beam}$
           &  $\phantom{-}$0.801    &  $\phantom{-}$0.798  
& $\phantom{-}$0.735   \\ \hline
$C_{\rm off}$ &  
           $\phantom{-}$0.808    &  $\phantom{-}$0.804 & 
$\phantom{-}$0.740   \\ \hline \hline
 \multicolumn{4}{|c|}{LHC} \\ \hline
$C_{\rm hel}$
          & 0.322    &  0.315  & 0.332  \\ \hline
\end{tabular}
\end{center}
\end{table}

In all results above we used $\mu=m_t=175$ GeV. A variation of
the scale $\mu$ between $m_t/2$ and $2m_t$  changes the central
results for $C$ at $\mu=m_t$ by roughly $\pm 5\%$.
Varying $m_t$ from $170$ to $180$ GeV changes the 
results for $C$ at the Tevatron by less than $5\%$, 
while for the LHC,  $C_{\rm hel}$ changes by less than a percent.  

\section{Conclusions}
\label{concl}

In summary, $t\bar{t}$ spin correlations are large effects that can
be studied at the Tevatron and the LHC by measuring double angular 
distributions both in the dilepton, single lepton and all-hadronic 
decay channels. The QCD corrections to these distributions are
of the order of 10 to 30$\%$. 
Spin correlations are suited to analyse in detail top quark
interactions, search for new effects, and may help to constrain the
parton content of the proton.  

\section*{Acknowledgments}
The results presented in this talk have been obtained in collaboration
with W. Bernreuther, Z.G. Si and P. Uwer.

\end{document}